\def\apj{{ApJ}}                 % Astrophysical Journal
\def\apjl{{ApJ}}                % Astrophysical Journal, Letters
\def\aap{{A\&A}}                % Astronomy and Astrophysics
\def\iaucirc{{IAU~Circ.}}       % IAU Cirulars
\def \estec  	{Astrophysics Division, Research and Science Support
		Department of ESA, ESTEC, Postbus 299, NL-2200 AG
		Noordwijk, The Netherlands}

\def \cesr	{Centre d'Etude Spatiale des Rayonnements, CNRS/UPS, 
		9 Av. du Colonel Roche, 31028 Toulouse Cedex 4, France}

\def \itesre	{Istituto Tecnologie e Studio Radiazioni Extraterrestri, 
		CNR, Via Gobetti 101, 40129 Bologna, Italy}

\def \vilspa	{XMM-Newton Science Operation Center, Vilspa, Spain}

\def\countsec{\hbox{counts s$^{-1}$}}

\def \rsun {\ifmmode$R$_{\odot}\else R$_{\odot}$\fi}
\def \nh {N${\rm _H}$}
\def \hcm {\hbox {\ifmmode $ atoms cm$^{-2}\else atoms cm$^{-2}$\fi}}

\def\approxgt{\mathrel{\hbox{\rlap{\lower.55ex \hbox {$\sim$}}
        \kern-.3em \raise.4ex \hbox{$>$}}}}
\def\approxlt{\mathrel{\hbox{\rlap{\lower.55ex \hbox {$\sim$}}
        \kern-.3em \raise.4ex \hbox{$<$}}}}

\newcommand {\sax} {{BeppoSAX}}

\def\arcsec{\hbox{$^{\prime\prime}$}}
\newcommand {\ergs} {erg~s$^{-1}$}
\newcommand {\ergcms} {erg~cm$^{-2}$~s$^{-1}$}
\newcommand {\chisq} {$\chi ^{2}$}
\newcommand {\rchisq} {$\chi_{\nu} ^{2}$}

\newcommand {\phind} {$\alpha$}
\newcommand {\prej} {P$_{\rm rej}$}
\newcommand {\ttnh} {$\times~$10$^{22}$~atom~cm$^{-2}$}

\def \src {XTE\,J0421+56}

\documentclass[a4paper]{VisionStyle}
\usepackage{epsfig}

\begin{document}

\title{Strongly absorbed quiescent X-ray emission from the X-ray
transient \src}

\author{L.\,Boirin\inst{1, 2} \and A.N.\,Parmar\inst{1} \and D.\,Lumb\inst{1} \and M.\,Orlandini\inst{3} \and N.\,Schartel\inst{4}}

\institute{
\estec
\and
\cesr
\and
\itesre
\and
\vilspa}

\maketitle 

\begin{abstract}

We have observed the soft X-ray transient \linebreak[4] \src\ in
quiescence with XMM-Newton. The observed spectrum is highly unusual
being dominated by a broad feature at 6.5~keV and can be modeled by a
strongly absorbed continuum. The spectra of X-ray transients observed
so far are normally modeled using Advection Dominated Accretion Flow
models, black-bodies, power-laws, or by the thermal emission from a
neutron star surface. The strongly absorbed X-ray emission of \src\
could result from the compact object being embedded within the dense
circumstellar wind emitted from the supergiant B[e] companion star.

\end{abstract}

\section{Introduction}
  
 \src\ was discovered by the All-Sky Monitor onboard RXTE as a soft
X-ray transient during its 1998 April outburst
(\cite{lboirin-C1:0421_smith98iau}). This outburst, one of the
shortest ever detected from a transient ($\sim$15 days) was observed
by CGRO (\cite{lboirin-C1:0421_paciesas98iau}), RXTE
(\cite{lboirin-C1:0421_belloni99apj}), ASCA
(\cite{lboirin-C1:0421_ueda98apjl}) and
\sax\ (\cite{lboirin-C1:0421_frontera98aa,lboirin-C1:0421_orr98aa}). The outburst spectra
are complex and can not be fit by any of the models usually applied to
the soft X-ray transients.

\src\ was also observed in quiescence by \sax\ (\cite{lboirin-C1:0421_orlandini00aa,lboirin-C1:0421_parmar00aa}). In 1998 September, the source was soft
with low absorption. In 1999 September, the source, in a sort of
re-flare, had hardened and brightened and became strongly absorbed
(\nh\ of (40~$\pm$10)~\ttnh). The 1-10 keV unabsorbed fluxes were
4.5~$\times$~10$^{-13}$ and 7.6~$\times$~10$^{-12}$ \ergcms\ in 1998
and 1999, respectively. The origin of X-ray emission from \src\
remains uncertain.

The optical counterpart of \src\ is CI~Cam, also known as MWC~84.
This is a supergiant B[e] star (\cite{lboirin-C1:clark99aa}, 2000;
\cite{lboirin-C1:0421_robinson01apj}). This makes \src\ the first high
mass X-ray binary (HMXB) with such a companion.

\nocite{lboirin-C1:0421_clark00aa}

Recently,
\cite*{lboirin-C1:0421_robinson01apj} estimated that the
distance to the source was greater than previously thought, making
\src\ among the most luminous transients. The  2-25~keV luminosity
at the peak of the outburst is 3.0~$\times$10$^{38}$~\ergs, assuming
the revised distance of 5~kpc.

\begin{figure}

\centerline{\includegraphics[width=0.45\textwidth]{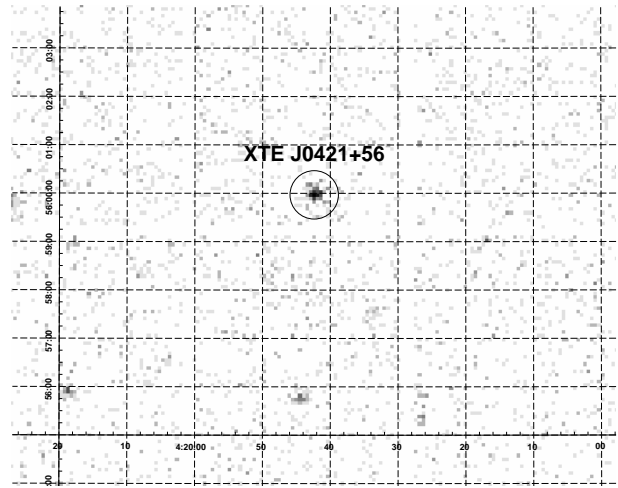}}

\caption{EPIC PN image of the region of sky containing \src. The radius of the circle showing the source position  is 30\arcsec.}

\label{fig:image}

\end{figure}

 Furthermore, \cite*{lboirin-C1:0421_robinson01apj} suggest that the
compact object is traveling in a wide orbit through the dense
spherical circumstellar wind emitted from the supergiant B[e]
star. Within this picture, the outburst would be caused by the same
disk instability mechanism responsible for the outbursts in X-ray
novae. It would thus differ from the outbursts observed in Be HMXB
that recur at multiples of the orbital period, probably when the
compact object crosses the equatorial wind from the Be star.

  \cite*{lboirin-C1:0421_robinson01apj} also favor a black hole as the compact
object, as infered from the absence of neutron star signatures (X-ray
bursts, periodic X-ray pulsations) and from the large ratio of peak to
quiescent luminosity.

\begin{figure}[!ht]

\centerline{\includegraphics[width=0.3\textwidth,angle=-90]{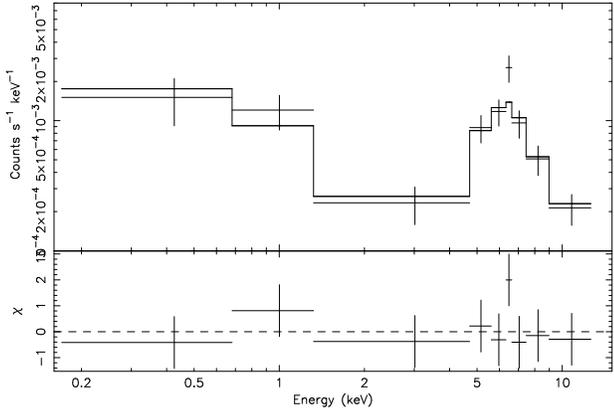}}

\caption{The PN spectrum of \src\ in quiescence (data and folded model). 
The line is the best-fit using the partialy absorbed power-law
model. The lower panel shows the residuals from the fit in terms of
sigmas.}

\label{fig:spectrum_count}
\end{figure}

%------------------------------------
%	Observations and data  analysis 
%------------------------------------

\section{XMM-Newton observation and data analysis}

 \src\ was observed in quiescence by \linebreak[4] XMM-Newton on 2001
August 19.  We present the spectral analysis of the EPIC PN data in
the 0.2-12 keV energy band where the source is detected.  The analysis
was carried out using the XMM-Newton Science Analysis Software. The
exposure is $\sim$~7~h after filtering out high background level
periods and selecting single pixel events.

Source counts were extracted from a circular region of 30\arcsec\
radius centered on \src\ (Figure~\ref{fig:image}). Background counts
were obtained from a circular region of 90\arcsec\ radius offset from
the source position.

The background count rate represents $\sim$37\% of the source count
rate (in the 0.2-12 keV band).  The net (source minus background)
count rate has a mean value of \linebreak[4] 0.0071$\pm$0.0007
\countsec.

In order to ensure applicability of the \chisq\ statistic with so few
counts, we rebinned the extracted spectrum such that at least 20 net
counts per bin were present. The obtained spectrum has 9 significant
bins (Figures~\ref{fig:spectrum_count} and \ref{fig:spectrum_photon}).
All spectral uncertainties are given at 90\% confidence.

%All spectral uncertainties and upper-limits are given at 90\% confidence. 

\section{The X-ray spectrum}

\begin{figure}[!t]

\centerline{\includegraphics[width=0.3\textwidth,angle=-90]{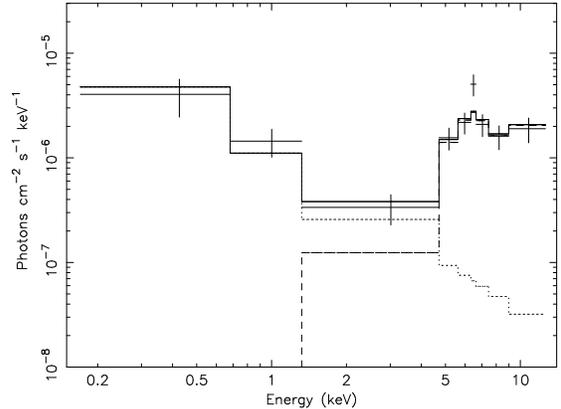}}

\caption{The PN unfolded spectrum of \src\ in quiescence. The line is the best-fit using the partialy absorbed power-law model. The dashed line shows the contribution of the absorbed and covered component to the total model. The doted line shows the contribution of the unabsorbed and uncovered component.}

\label{fig:spectrum_photon}
\end{figure}

 The spectrum of \src\ (Figures~\ref{fig:spectrum_count} and
\ref{fig:spectrum_photon}) is clearly dominated by a broad feature
peaking at 6.5 keV.

Considering that the feature around 6.5~keV could be due to an
emission from Fe, we first tried a model consisting of a power-law to
account for the low energy component and a Gaussian to account for the
6.5 keV feature.  This model does not provide an acceptable fit to the
data: the reduced \chisq\ (\rchisq) is 3.8 for 4 degrees of freedom
(dof). A similar result is obtained when including photo-electric
absorption. A model with a power-law and two Gaussians does not
provide a good fit either.

We then adopted a different fitting approach, considering the 6.5 keV
feature not as an emission line, but as a highly absorbed continuum
component.  This approach is consistent with the one adopted to fit
the \sax\ spectrum of \src\ in quiescence during the 1999 September
observation (\cite{lboirin-C1:0421_parmar00aa}).

A model consisting of a power-law (photon index, \phind, of
1.4$\pm$0.4) plus an absorbed power-law (\nh\ of \linebreak
(72$\pm$37)~\ttnh, \phind\ of 2$\pm$1) was tried. This gives a
\rchisq\ of 1.35 for 4 dof.  An F-test indicates that adding a
Gaussian at 6.5 keV does not improve the fit: the probability \prej\
of rejecting the hypothesis that the fit is better including the
additional Gaussian is 98\%.

A better fit (\rchisq\ of 1.07 for 5 dof) is obtained using a partialy
absorbed power-law model, with an \nh\ of (61$_{-20}^{+27}$)~\ttnh,
\phind\ of 1.5$\pm$0.5 and a partial covering fraction of
0.99$_{-0.010}^{+0.007}$.  This model is shown in
Figures~\ref{fig:spectrum_count} and \ref{fig:spectrum_photon}. Adding
a Gaussian at 6.5 keV does not significantly improve the fit (\prej\
of 69\%).

In Figure~\ref{fig:spectrum_photon}, the contributions of the absorbed
and covered component (dashed line) and of the unabsorbed and
uncovered component (doted line) to the total model are shown
separetely. The absorbed and covered component clearly dominates the
emission above $\sim$5~keV whereas the unabsorbed and uncovered
component dominates the emission below $\sim$5~keV.

The 1-10 keV absorbed and unabsorbed (derived by setting \nh\ to 0)
fluxes are 1.3 and 9.3~$\times$~10$^{-13}$~\ergcms\ respectively,
using the partialy absorbed power-law model. This corresponds to a
luminosity of \linebreak 2.8~$\times$~10$^{33}$~\ergs, assuming a
distance of 5~kpc.

%------------------------------------
%	Discussion
%------------------------------------

\section{Discussion}

 We have shown that the quiescent X-ray spectrum of \linebreak \src\
during the XMM-Newton observation was best modeled by a strongly
absorbed power-law (\nh\ of (61$_{-20}^{+27}$)~\ttnh).

Such a spectrum is unique among soft X-ray transients, and adds to the
growing list of very peculiar properties observed from \src.

Although we note that only simple models could be tested because of
the small number of bins in the spectrum, this model is very
attractive because it indicates the presence of large amounts of
absorbing material within the system itself. Thus, this confirms the
recently drawn picture of the compact object being embedded within a
very dense wind emitted from the B[e] companion star
(\cite{lboirin-C1:0421_robinson01apj}).

Furthermore, it suggests that the accretion onto the compact object is
still occuring at a substantial rate in order to explain the observed
quiescent luminosity. This is also consistent with a compact object
continuously travelling through an extended wind and not only crossing
an equatorial wind during short intervals as in Be systems.

\section*{Acknowledgements}

Based on observations obtained with XMM-Newton, an ESA science mission
with instruments and contributions directly funded by ESA member
states and the USA \linebreak (NASA).  L.~Boirin acknowledges an ESA
Fellowship.

%------------------------------------
%	References
%------------------------------------


\begin{thebibliography}{}

\bibitem[\protect\astroncite{{Belloni} et~al.}{1999}]{lboirin-C1:0421_belloni99apj}
{Belloni} T., {Dieters} S., {van den Ancker} M.~E. et~al.,  1999,
\newblock {\apj} {527}, 345


\bibitem[\protect\astroncite{{Clark} et~al.}{1999}]{lboirin-C1:clark99aa}
{Clark} J.~S., {Steele} I.~A., {Fender} R.~P. et~al., 1999,
\newblock {\aap} {348}, 888

\bibitem[\protect\astroncite{{Clark} et~al.}{2000}]{lboirin-C1:0421_clark00aa}
{Clark} J.~S., {Miroshnichenko} A.~S., {Larionov} V.~M. et~al.,  2000,
\newblock {\aap} {356}, 50


\bibitem[\protect\astroncite{{Frontera} et~al.}{1998}]{lboirin-C1:0421_frontera98aa}
{Frontera} F., {Orlandini} M., {Amati} L. et~al., 1998,
\newblock {\aap} {339}, L69


\bibitem[\protect\astroncite{{Orlandini} et~al.}{2000}]{lboirin-C1:0421_orlandini00aa}
{Orlandini} M., {Parmar} A.~N., {Frontera} F. et~al.,  2000,
\newblock {\aap} {356}, 163

\bibitem[\protect\astroncite{{Orr} et~al.}{1998}]{lboirin-C1:0421_orr98aa}
{Orr} A., {Parmar} A.~N., {Orlandini} M. et~al.,  1998,
\newblock {\aap} {340}, L19


\bibitem[\protect\astroncite{{Paciesas} \&
  {Fishman}}{1998}]{lboirin-C1:0421_paciesas98iau} {Paciesas} W. and
  {Fishman} G., 1998,
\newblock {\iaucirc} {6856}

\bibitem[\protect\astroncite{{Parmar} et~al.}{2000}]{lboirin-C1:0421_parmar00aa}
{Parmar} A.~N., {Belloni} T., {Orlandini} M. et~al., 2000,
\newblock {\aap} {360}, L31

\bibitem[\protect\astroncite{{Robinson} et~al.}{2001}]{lboirin-C1:0421_robinson01apj}
{Robinson} E.~L., {Ivans} I.~I., and {Welsh} W.~F., 2001,
\newblock {\apj},
\newblock accepted for publication, astroph/0110110


\bibitem[\protect\astroncite{{Smith} et~al.}{1998}]{lboirin-C1:0421_smith98iau}
{Smith} D., {Remillard} R., {Swank} J. et~al., 1998, \newblock
{\iaucirc} {6855}
  
\bibitem[\protect\astroncite{{Ueda} et~al.}{1998}]{lboirin-C1:0421_ueda98apjl}
{Ueda} Y., {Ishida} M., {Inoue} H. et~al., 1998,
\newblock {\apjl} {508}, L167


  \end{thebibliography}
 \end{document}